# COOL DWARFS IN WIDE MULTIPLE SYSTEMS

## PAPER 6: A CURIOUS QUINTUPLE SYSTEM OF A COMPACT SUN-LIKE TRIPLE AND A CLOSE PAIR OF AN M DWARF AND A VERY COOL WHITE DWARF AT A WIDE SEPARATION

*By Rodrigo González-Peinado,*
*Departamento de Física de la Tierra, Astronomía y Astrofísica & UPARCOS,*
*Universidad Complutense de Madrid, Spain*

*José A. Caballero*
*Centro de Astrobiología (CSIC-INTA), Madrid, Spain*

*David Montes*
*Departamento de Física de la Tierra, Astronomía y Astrofísica & UPARCOS,*
*Universidad Complutense de Madrid, Spain*

*and Carlos Cifuentes*
*Centro de Astrobiología (CSIC-INTA), Madrid, Spain*

The system WDS 16329+0315 is an old, nearby quintuple physical system in the thick Galactic disc formed by a close-resolved, triple primary of solar metallicity, namely HD 149162, and a very wide, common proper motion, secondary pair, formed by the mid-M dwarf G-17-23 and the white dwarf LSPM J1633+0311S. We present an exhaustive astrometric and photometric data compilation of the system, including *Gaia* DR2 parallaxes and proper motions, and the first analysis of the nature of the faintest component. LSPM J1633+0311S (HD 149162 C) is a very cool white dwarf with an effective temperature of only about 5500 K, near the coolest end of the grid of theoretical models.

*The WDS 16329+0315 system*

HD 149162 is the brightest member of the multiple stellar system WDS 16329+0315. The star was first classified in 1947 as a dK0 single-line spectroscopic binary[1]. The first careful study of the spectroscopic binary, hereafter Aa,Ab, was carried out by Johnson & Mayor[2]. They calculated an orbital period $P$ = 225.7±1.1 d and a moderate eccentricity $e$ = 0.282±0.002, and kinematically classified it as an old-disc K1V-type binary. They computed the mass of the primary, Aa, at 0.8-0.9 $M_\odot$ from the spectroscopic mass function and at 0.76 $M_\odot$ from Geneva photometry, and of the "secondary", Ab, at over 1 $M_\odot$. To explain this contradictory result and the significant X-ray, ultraviolet and Ca II H&K emission, they proposed the secondary to be, in turn, a neutron star, a white dwarf-M dwarf tight binary, or a pair of emission-line M dwarfs.

Subsequent kinematic studies and orbital solutions also supported the idea of the system being part of the old disc, and the Ab component being more massive than Aa[3, 4, 5]. For example, HD 149162 was one of the 235 stars for which Lindegren *et al.*[4] published orbital solutions from *Hipparcos* data. Next, Horch *et al.*[6] used the Gemini North telescope and the Differential Speckle Survey Instrument for studying the mass-luminosity relation in metal-poor stars. As part of their analysis, they resolved the components Aa and Ab, which were separated by only $\rho$ = 0.007-0.020 arcsec. All the orbital solutions published in the literature for the pair Aa,Ab are listed in Table I. Both astrometric and spectroscopic analyses point to an orbital period of about 0.62 a with a moderate eccentricity of about 0.31.

Horch *et al.*[6] also discovered a third fainter source, hereafter Ac, at $\rho \sim$ 0.284 arcsec from Aa, and measured magnitude differences at two narrow filters centred on 692 nm and 880 nm. The observed differences agreed with Ab and Ac being k6V and m5V main-sequence stars, respectively (a lower case letter in spectral type stands for a determination based on photometry only). Strikingly, given its angular separation, Ac cannot be one of the M dwarf companion candidates proposed by Johnson & Mayor[2]. The *Washington Double Star Catalogue (WDS*[7]*)* tabulates the triple system with the discoverer code DSG 7.

There has been a long discussion in the literature on the actual iron abundance (the most used proxy of metallicity) of HD 149162 Aa, which is the only component visible in spectroscopic analyses. The iron abundance has been reported in the wide range from –1.39 to –0.04[2, 3, 8], which made Lindegren *et al.*[4] to include it in their sample. However, Montes *et al.*[9] settled the issue and measured a solar metallicity ([Fe/H] = –0.01±0.04). Montes *et al.*[9] also derived accurate Galactocentric space velocities, *UVW*, for Aa, and assigned it to the Galactic thick disc, in accordance with previous classifications and the lack of identified activity (i.e. youth) features, such as H$\alpha$ or X-ray emission. This points towards a relatively old systemic age.

Lépine & Bongiorno[10] found a wide common proper motion companion to HD 149162, at an angular separation of $\rho$ = 252.0 arcsec and position angle $\theta$ = 138 deg, namely G 17-23, hereafter B. This pair is catalogued by WDS as LEP 98. In the second half of the 20$^{th}$ century, this M3.0 V star[11] was thought to be isolated[12]. As a relatively bright intermediate M dwarf, B was a potential target for exoplanet search by CARMENES[11, 13], which originally attracted our attention.

Besides, Lépine & Shara[14] tabulated a faint source at 6.30 arcsec to the southeast of B with exactly the same proper motions. This component, namely LSPM J1633+0331S, hereafter C, has been poorly investigated: before Montes *et al.*[9], only coordinates, proper motions and *J*, *H*, and $K_s$ magnitudes had been reported[13]. The pair B,C is catalogued by WDS as DAM 649.

Table II summarises the known parameters of the quintuple system. Fig. 1 shows an image of the system as visualised with Aladin[15].

*Data compilation and analysis*

First of all, we compiled equatorial coordinates (right ascension and declination) for the Aa, B and C components with their astrometric epochs, from the following catalogues: AC2000.2[16], USNO-A2[17], GSC2.3[18], 2MASS[19], CMC15[20],

AllWISE[21], and Gaia-DR1[22]. In addition, we measured coordinates for the B and C components in eight digitised photographic plates downloaded from the Digital Sky Survey and SuperCOSMOS[23]. The used plates had exposure times between 12 and 95 min, and different emulsion-filter combinations from the blue to the near infrared. With these coordinates, we derived precise angular separations and position angles between Aa and B components and between B and C components for each astrometric epoch with the Virtual Observatory tool TOPCAT[24]. We list these measurements, along with average values, in Table III.

We also calculated proper motions for the three stars with simple linear regressions in the epoch-α and epoch-δ planes[25]. In total, we used seven different astrometric epochs for the Aa component with a time baseline of 105 years. For the B and C components, the time baseline of the astrometric follow-up was shorter, 65 years, but contained more data points: 13 and 10, respectively. The three stars share the same proper motion within uncertainties, and are very similar to the new values in the *Gaia* Second Data Release (*Gaia* DR2[26]) listed in Table II and to pre-*Gaia* DR2 values from the literature, as shown in Table IV.

Finally, we collected magnitudes for the three components resolved by all-sky surveys in 11 different wavelengths from 490 nm to 2300 nm. The compiled magnitudes are *g*, *r*, *i*, *z*, and *y* from Pan-STARRS[27], $B_P$, *G* and $R_P$ from *Gaia*-DR2[26], and *J*, *H*, and $K_s$ from 2MASS[18]. These values are presented in Table II. Only for component B, we also collected *B* and *V* magnitudes from APASS9[28] and *W1-4* magnitudes from AllWISE[21], which amounted 17 photometric pass-bands for the M dwarf.

*Results, discussion and conclusions*

As expected from the common proper motion, the three stars resolved by *Gaia* (A, B and C) also have similar parallactic distances (Table II). However, the close multiplicity of Aa,Ab,Ac affects the *Gaia* measures, at the triple star has a larger parallax uncertainty than B and C, in spite of being several magnitudes brighter (but fainter than the *Gaia* bright end at *G* ~ 6 mag). Therefore, we assumed that the multiple system is located at *d* = 45.48±0.16 pc, which is the distance obtained from the weighted mean of the parallaxes of B and C only. Compare this value with *d* = 46.3±1.9 pc, which was tabulated for Aa,Ab,Ac by the Tycho-*Gaia* Astrometric Catalogue (TGAS[22]).

The average angular separations ρ and position angles θ for the pairs Aa,B and B,C listed in Table III, ρ = 252.35±0.21 arcsec and θ = 138.28±0.04 deg, and ρ = 6.01±0.11 arcsec and θ = 135.19±0.07 deg, respectively, match (and slightly improve) the ones provided by WDS. Together with the assumed heliocentric distance to the system, these angular separations and those measured by Horch *et al.*[6] with speckle translate into projected physical separations *s* = 13.02±0.06 au (Aa-Ac), *s* = 11570±40 au (Aa,B), *s* = 276±5 au (B,C). With the orbital parameters in Table I, the true physical separation between Aa and Ab varies between *r* = 0.47±0.06 au at periastron and *r* = 0.89±0.12 au at apoastron. All in all, WDS 16329+0315 is a hierarchical quintuple system with different physical separations ranging in six orders of magnitude (from $10^{-1}$ au to $10^{5}$ au).

Interestingly, component C (LSPM J1633+0311S), which has never been spectroscopically investigated, is 4.4 mag fainter in *G* band than component B, which is an M3.0 dwarf[11]. Since they are located at the same distance, this would make C to be a very red late M dwarf. However, C is *bluer* than B in all colours. This is illustrated by the colour-magnitude diagram shown in Fig. 2. Because of its faint absolute magnitude and blue colours, we considered component C as a white dwarf or hot subdwarf candidate. Montes *et al.*[9] assigned a "D:" (white dwarf) spectral type to LSPM J1633+0311S based on information provided in *this* work.

With the photometry compiled in Table II, the widely-used Koester *et al.*[29] theoretical models of white dwarfs and the Virtual Observatory Spectral Energy Distribution Analyzer (VOSA[30]), we derived an effective temperature of 5500±50 K, a surface gravity log $g$ = 9.5±0.5 and a bolometric luminosity $L$ = (122±8) $10^{-6}$ $L_{sol}$ for C. The low effective temperature, near the coolest end of the Koester *et al.*[29] grid of models at 5000 K, high surface gravity and small luminosity agree with what is expected for a very cool white dwarf[31, 32, 33, 34] and the most recent findings with *Gaia-DR1*[35, 36,] and *Gaia*-DR2[36, 37, 38]. The low temperature is, however, not compatible with C being a hot subdwarf. Although we do not know the nature of the stellar progenitor or the cooling evolution of C, its cool temperature supports the old systemic age stated above. Besides, we proceed similarly with B, but with BT-Settl CIFIST[40] models of low-mass stars and additional optical APASS9 and mid-infrared *WISE* photometry; again, the derived temperature, gravity and luminosity, $L$ = (164±2) $10^{-4}$ $L_{sol}$, match the typical parameters of an M3.0V dwarf[41].

Finally, we estimated the temporal scales of variation in the system. Components Aa (K1V) and Ab (k6V), with masses of 0.77 $M_{\odot}$ and 0.64 $M_{\odot}$ according to Horch *et al.*[6], orbit around their common centre of gravity every 0.619 a (Table I). For Ac (m5V) and B (M3.0V), we assumed masses of 0.18 and 0.40 $M_{\odot}$ from typical masses for intermediate M dwarfs[42, 43, 44], and from the *J*-band absolute magnitude, parallactic distance to the system, and the BT-Settl Lyon models[40] for ages between 1 and 10 Ga. For component C, we adopted the typical mass for a white dwarf at 0.54 $M_{\odot}$[31]. Next, we computed *reduced orbital periods, P\**, from the third Kepler's law of planetary motion $G (M1+M2) P^2 = 4\pi\ a^3$, where the semi-major axis *a* was replaced by the mean projected physical separation *s*, as in Caballero[25]. Resulting periods were 37 a (Aa,Ab-Ac), 4700 a (B,C) and 810,000 a (Aa,Ab,Ac-B,C). The indirect effect of Ac on the spectrum of Aa, or even the relative proper motion of Ac with respect to Aa and Ab, could be measured within a few years, which would shed light on the nature of the low-mass star companions to the K1V star. However, we need low- and high- resolution spectroscopy for studying in further detail the new very cool white dwarf LSPM J1633+0311S, which we baptise here as HD 149162 C.

*Acknowledgements*

We thank F. Jiménez-Esteban for helpful discussion on *Gaia* DR2 white dwarfs. This research made use of the SIMBAD database and VizieR catalogue access tool, operated at Centre de Données astronomiques de Strasbourg, France, the Spanish Virtual Observatory, the NASA's Astrophysics Data System, and the Washington Double Star Catalog maintained at the U.S. Naval Observatory. Financial support was provided by the Spanish MICINN under grants AYA2016-79425-C3-1/2-P, and the Conserjería de Educación, Juventud y Deporte de la Comunidad de Madrid and the Fondo Social Europeo y la Iniciativa de Empleo Juvenil (YEI) under grant PEJD-2016/TIC-2347.

[Notes to the Editor: we use the symbol 'a' (SI, IAU: *annus*) for the unit 'year'.]

| TABLE I |
| --- |
| Orbital solution parameters of pair WDS 16329+0315 Aa,Ab |

*From spectroscopy*

| Ref. | $P$ [d] | $T_0$ [MJD] | $e$ | $\omega$ [deg] | $a \sin i$ [Gm] | $f(M)$ [M$_\odot$] |
| --- | --- | --- | --- | --- | --- | --- |
| 2 | 225.70±0.10 | 45282.1±0.3 | 0.282±0.002 | 17.6±0.5 | 71.10±0.30 | 0.289±0.003 |
| 3 | 226.30±1.30 | 47319.1±1.4 | 0.324±0.012 | 19.5±2.3 | 69.34±0.94 | 0.260±0.010 |
| 5 | 226.08±0.18 | 47319.3±1.1 | 0.3114±0.0087 | 20.4±1.7 | 69.76±0.78 | 0.2647±0.0089 |

*From astrometry*

| Ref. | $P$ [d] | $T_0$ [MJD] | $e$ | $\omega$ [deg] | $a$ [mas] | $i$ [deg] |
| --- | --- | --- | --- | --- | --- | --- |
| 4 | 225.70 | 45282.1 | 0.282 | 17.60 | 9.87±1.62 | 109.47±15.28 |
| 6 | 226.08 | 47319.5 | 0.3114 | 203.62 | 14.80±2.00 | 112±26 |

| TABLE II | | | | |
| --- | --- | --- | --- | --- |
| Fundamental parameters of WDS 16329+0315 Aa,Ab,Ac, B and C | | | | |
| Component | Aa,Ab,Ac | B | C | Ref. |
| Name | HD 149162 | G 17-23 | LSPM J1633+0311S | Simbad |
| Alt. name | BD+03 3215 | NLTT 43046 | … | |
| $\alpha$ [h:m:s] | 16:32:51.631 | 16:33:02.799 | 16:33:03.088 | *Gaia*-DR2[26] |
| $\delta$ [d:m:s] | +03:14:45.64 | +03:11:37.28 | +03:11:32.63 | |
| $\mu_\alpha \cos\delta$ [mas/a] | −374.95±0.52 | −368.92±0.08 | −369.28±0.24 | |
| $\mu_\delta$ [mas/a] | −180.81±0.44 | −186.07±0.06 | −189.64±0.16 | |
| $d$ [pc] | 41.65±0.62 | 45.38±0.10 | 45.57±0.30 | |
| $B_P$ [mag] | 9.0635±0.0011 | 14.8338±0.0016 | 18.1523±0.0082 | |
| $G$ [mag] | 8.5731±0.0003 | 13.4083±0.0003 | 17.8064±0.0013 | |
| $R_P$ [mag] | 7.9516±0.0013 | 12.2391±0.0012 | 17.1870±0.0149 | |
| $g$ [mag] | … | 15.171±0.005 | 18.185±0.008 | Pan-STARRS[27] |
| $r$ [mag] | … | 13.978±0.003 | 17.756±0.005 | |
| $i$ [mag] | 8.393±0.108 | 12.706±0.012 | 17.594±0.012 | |
| $z$ [mag] | 8.386±0.001 | 12.144±0.014 | 17.524±0.008 | |
| $y$ [mag] | 8.236±0.115 | 11.830±0.005 | 17.398±0.018 | |
| $J$ [mag] | 7.159±0.024 | 10.625±0.026 | 16.314±0.279 | 2MASS[19] |
| $H$ [mag] | 6.700±0.055 | 10.011±0.023 | 15.918±0.352 | |
| $K_s$ [mag] | 6.561±0.018 | 9.775±0.021 | … | |
| Sp. type | K1V+k6V+m5V | M3.0V | D: | 2, 6, 11, 9 |

| $T_{eff}$ [K] | 5252±53 | 3400±50 | 5500±50 | [9], this work |
|---|---|---|---|---|
| log $g$ | 4.33±0.13 | 4.5 (fixed) | 9.5±0.5 | |
| [Fe/H] | −0.01±0.04 | 0.00 (fixed) | ... | |
| $V_r$ [km/s] | −51.33±0.15 | ... | ... | |
| $U$ [km/s] | −41.79±0.12 | ... | ... | |
| $V$ [km/s] | −94.4±3.3 | ... | ... | |
| $W$ [km/s] | +14.8±1.7 | ... | ... | |
| Population | Thick disc | ... | ... | |
| $M$ [M$_\odot$] | (0.77+0.64+0.15)±0.02 | 0.40 | 0.54 | [6], this work |

| TABLE III | | | | |
|---|---|---|---|---|
| *Astrometric follow-up of pairs WDS 16329+0315 Aa-B and B-C* | | | | |
| *Pair* | *Epoch* | $\rho$ [arcsec] | $\theta$ [arcsec] | *Origin* |
| Aa,B (LEP 79) | 1950.294 | 252.74±0.10 | 138.47±0.01 | USNO-A2[17] |
| | 1993.530 | 253.16±0.14 | 137.78±0.01 | GSC2.3[18] |
| | 1999.938 | 252.16±0.46 | 138.26±0.09 | CMC15[20] |
| | 2000.350 | 252.01±0.06 | 138.39±0.01 | 2MASS[19] |
| | 2010.556 | 252.01±0.04 | 138.40±0.01 | AllWISE[21] |
| | 2015.000 | 252.03±0.43 | 138.39±0.01 | *Gaia*-DR1[22] |
| | *Average* | 252.35±0.21 | 138.28±0.04 | |
| | | | | |
| B,C (DAM 649) | 1950.294 | 5.88±0.13 | 127.37±0.25 | POSSI-Red XE565 |
| | 1983.452 | 5.99±0.05 | 133.50±0.02 | Quick-V-Northern N565 |
| | 1990.384 | 6.36±0.05 | 135.02±0.01 | POSSII-Blue XJ800 |
| | 1993.390 | 6.08±0.05 | 135.04±0.01 | POSSII-Red XP800 |
| | 1993.475 | 5.98±0.05 | 138.41±0.04 | POSSII-Blue XJ801 |
| | 1993.530 | 5.80±0.05 | 134.72±0.01 | POSSII-Red XP801 |
| | 1997.292 | 5.79±0.05 | 135.12±0.01 | POSSII-IR XI800 |
| | 1997.440 | 5.66±0.05 | 136.15±0.01 | POSSII-IR XI801 |

|   | 2000.350 | 6.26±0.20 | 138.51±0.17 | 2MASS[19] |
|---|---|---|---|---|
|   | 2015.000 | 6.39±0.31 | 137.44±0.18 | *Gaia*-DR1[22] |
|   | *Average* | 6.01±0.11 | 135.19±0.07 |   |

| TABLE IV | | | | | |
|---|---|---|---|---|---|
| *Calculated and pre-Gaia-DR2 proper motions of WDS 16329+0315 Aa,Ab,Ac, B and C* | | | | | |
| Component | $\mu_\alpha \cos\delta$ [mas/a] | $\mu_\delta$ [mas/a] | Epochs | $\mu_\alpha \cos\delta_{lit}$ [mas/a] | $\mu_{\delta lit}$ [mas/a] | Origin |
| Aa,Ab,Ac | −370.9±1.4 | −186.8±1.5 | 7 | −369.81±0.07 | −185.50±0.05 | TGAS[22] |
| B | −374.9±4.9 | −179.9±3.3 | 13 | −383.38±2.31 | −172.58±2.31 | HSOY[45] |
| C | −380.4±5.8 | −197.5±5.3 | 10 | −374 | −189 | [14] |

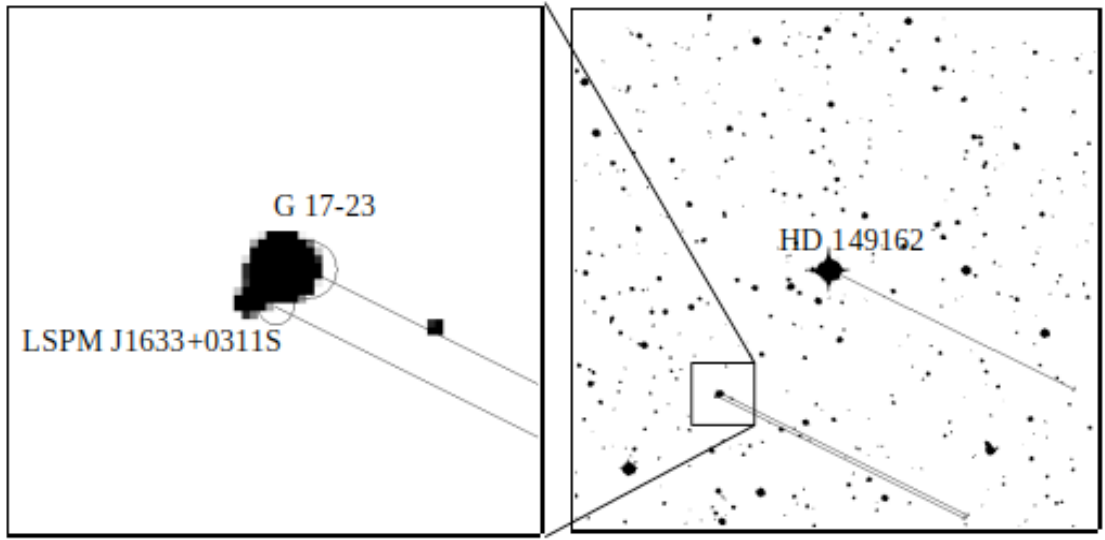

FIG. 1
*Inverted Digital Sky Survey $R_F$ photographic images of the system WDS 16329+0315 made with Aladin.* Left: *B and C components, approximately 2 x 2 arcmin². Right: Aa,Ab,Ac (unresolved) and B,C (unresolved), approximately 30 x 30 arcmin². Names are labeled. Vectors indicate Simbad proper motions. North up and East to the left.*

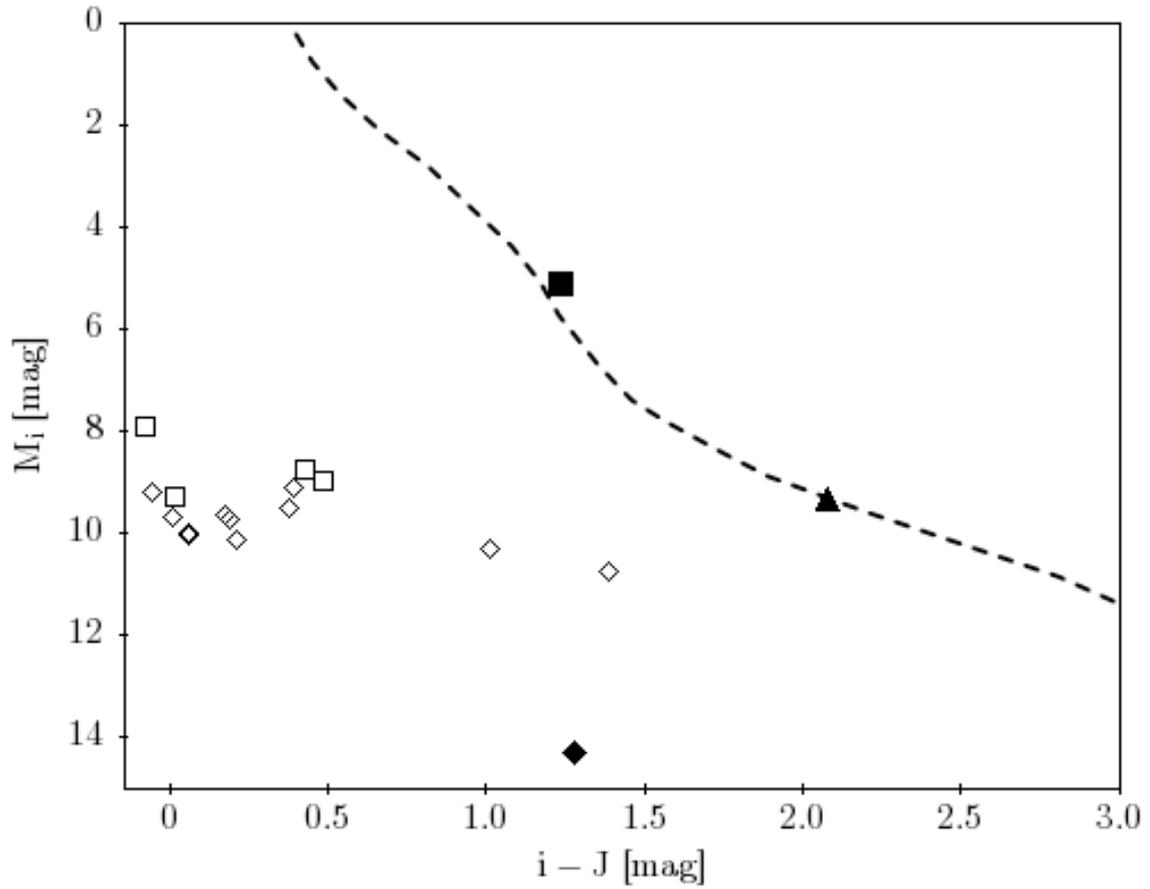

FIG. 2

*Optical-near infrared colour-magnitude diagram of the system WDS 16329+0315. Large black filled symbols: components Aa,Ab,Ac (square; slightly above the isochrone because of unresolved multiplicity), B (triangle) and C (rhomb). Small open symbols: the brightest white dwarfs (rhombs) and hot subdwarfs (squares) in the Tycho-2 catalogue[34]. Dashed line: 10 Ga-old BT-Settle CIFIST Lyon isochrone[45].*